# Science Priorities for the Extraction of the Solid MSR Samples from their Sample Tubes

**NASA-ESA Mars Rock Team**

Nicolas Dauphas, Sara S. Russell, David Beaty, Fiona Thiessen,

Jessica Barnes, Lydie Bonal, John Bridges, Thomas Bristow, John Eiler, Ludovic Ferrière, Teresa Fornaro, Jérôme Gattacceca, Beda Hoffman, Emmanuelle J. Javaux, Thorsten Kleine, Harry Y. McSween, Manika Prasad, Liz Rampe, Mariek Schmidt, Blair Schoene, Kirsten L. Siebach, Jennifer Stern, Nicolas Tosca.

**Background:** The *NASA-ESA Mars Rock Team* is an outgrowth of the MCSG1 team. It is composed of scientists with expertise in handling and analyses of both terrestrial and extraterrestrial samples, rock physics, and contamination mitigation. Two online meetings were organized in the Fall of 2022 where Oscar Rendon Perez (JPL) and Paulo Younse (JPL) described the engineering options for opening the tubes that will contain the samples returned from Mars' Jezero crater. This prompted discussions between the Rock Team members (during online meetings and through emails). The Rock Team leadership met online with the team focused on gas analysis (Gas Team) to understand their constraints and make sure that the solutions envisioned for headspace gas extraction would not compromise solid core retrieval. This report summarizes the consensus view of the Rock Team. It was written by the Rock Team leadership with input from all team members.

**Summary:** Preservation of the chemical and structural integrity of samples that will be brought back from Mars is paramount to achieving the scientific objectives of MSR. Given our knowledge of the nature of the samples retrieved at Jezero by Perseverance, at least two options need to be tested for opening the sample tubes: (1) One or two radial cuts at the end of the tube to slide the sample out. (2) Two radial cuts at the ends of the tube and two longitudinal cuts to lift the upper half of the tube and access the sample. Strategy 1 will likely minimize contamination but incurs the risk of affecting the physical integrity of weakly consolidated samples. Strategy 2 will be optimal for preserving the physical integrity of the samples but increases the risk of contamination and mishandling of the sample as more manipulations and additional equipment will be needed. A flexible approach to opening the sample tubes is therefore required, and several options need to be available, depending on the nature of the rock samples returned. Both opening strategies 1 and 2 may need to be available when the samples are returned to handle different sample types (*e.g.*, loosely bound sediments vs. indurated magmatic rocks). This question should be revisited after engineering tests are performed on analogue samples. The MSR sample tubes will have to be opened under stringent BSL4 conditions and this aspect needs to be integrated into the planning.

**Introduction:** NASA-ESA are planning to collect and transport from Mars to Earth a set of samples of martian materials for the purpose of scientific investigation (Kminek *et al.* 2022). The samples are currently collected by the Perseverance Rover (Farley and Stack, 2022) and consist of rocks, regolith, and at least one dedicated sample of atmospheric gas. In addition, for the rock and regolith samples, the process of sealing the sample tubes at the martian surface will result in the volume above the solid samples (referred to as the head space) being occupied by martian atmospheric gas. The samples will be contained within titanium sample tubes, which will be sealed at the martian surface with a compression-style cap.

The rocks sampled thus far by the Perseverance Rover comprise magmatic rocks like basalt and olivine cumulates that experienced various degrees of secondary water alteration, water-laid detrital sedimentary rocks that show various levels of induration, and unconsolidated Mars regolith that could contain grains from afar transported to the Jezero crater. Two main considerations weigh on the strategy that should be adopted for opening the samples:

(1) Important information is contained in the vertical successions and textural characteristics of layers in sediments, which can provide important clues for interpreting the depositional setting (Fig. 1). For example, in terrestrial lakes, vertical gradation in grain size can reflect the relative density of depositional and lacustrine fluids or gradations in organic matter content can reflect seasonal changes in biological productivity. Fine laminations can sometimes reflect the presence of microbial mats. The method used for opening the tubes must imperatively preserve those fine structures.

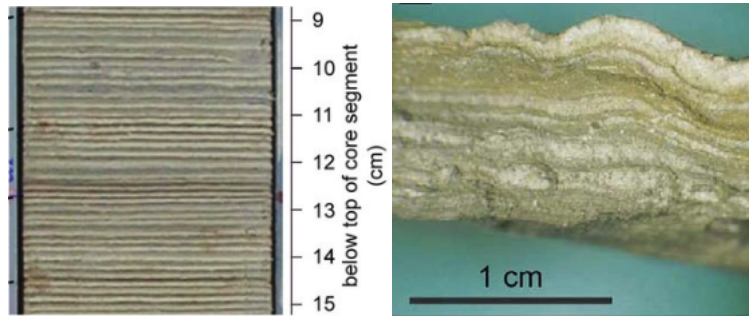

***Fig. 1.*** *Examples of possible fine-scale laminations in terrestrial environments (left; seasonal varves from Lake Belau, Northern Germany; Dörfler et al. 2012; right Microbially-Induced Sedimentary Structures-MISS in the middle neoproterozoic Chuar Group, Grand Canyon, Arizona; Bohacs and Junium 2007).*

(2) Some critical measurements are sensitive to contamination either from the tube, the apparatus used for cutting the tubes, or surrounding contaminants present in the isolator. Organic matter is of particular concern given the high stakes involved in any claim for the presence of any form of biotic or prebiotic chemistry on Mars. Inorganic trace element isotopes may provide dates on when Mars was habitable, and these are also prone to contamination.

Beginning in 2022, an engineering team was tasked with developing the processes needed to open the sample tubes and to extract the solid and gaseous samples. The engineering team was asked to develop engineering priorities associated with this process. Two science teams were asked to develop parallel science priorities: A group we call the "Gas Team" evaluated the priorities related to the science associated with all returned gaseous sample, and a second group called the "Rock Team" (the authors of this report) evaluated the priorities associated with solid materials contained within the sample tubes. Both the "Gas Team" and "Rock Team" work under the oversight of a third committee, the Mars Campaign Science Group (MCSG1).

The solid samples returned from the martian surface are certain to include sedimentary rocks (most important for the search for biosignatures), igneous rocks, and regolith, and they may also include other kinds of rocks, such as hydrothermal rocks, or impact breccia. The samples will be the basis for answering the main scientific questions of Mars Sample Return (iMOST, 2018).

The rock samples at Mars will all have been collected from various outcrops (or perhaps very large blocks of coherent rock). However, at least some of the rocks are relatively weak (i.e. a low compressive strength), and are vulnerable to fracturing during drilling and during several dynamic events associated with spacecraft operations during the return phase (most importantly, at Earth landing). It is anticipated that the mechanical state of each sample, as received in the laboratory on Earth, will be assessed by a method like computer tomography (CT) scanning prior to opening. The decision on how to open each sample tube can therefore be based on geological data from the field (collected by the M2020 science team), tests done on analogue samples, as well as the penetrative imaging data obtained on Earth.

The engineering team has proposed a 2-phase process for opening the sample tubes: First, puncture the tube in a way that will allow any gas present to be extracted and captured, then second, cut the metal of the tube in a way that would allow the solid materials to be removed. Regarding cutting the metal of the tubes, three primary mechanisms have been proposed (Fig. 2):

- A single radial cut to the end of the tube, so that the sample could be tipped out.
- A radial cut at each end of the tube, which would enable the sample to be pushed out from one end

- Two radial cuts and two longitudinal cuts, to reveal the whole sample during cutting.

An option frequently used on Earth to access core samples, for example used with deep sea drill cores, is to cut the core tube and the core together with something like a band saw. This is not an option for samples returned from Mars as this would have the effect of driving contamination from both the metallic core tube and band saw into the interior of the rock core.

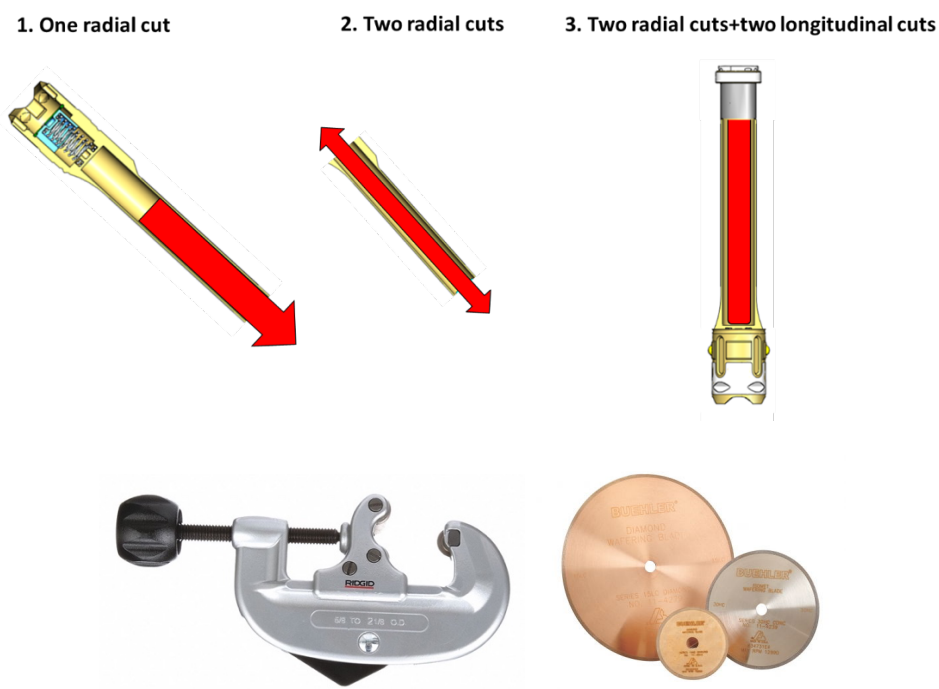

*Figure 2. Proposed protocols for opening the sample tubes. Drawings courtesy of Oscar Rendon Perez. In the one radial cut approach, a sharp hard metal wheel shears through the tube by slowly rotating and tightening it around the tube (bottom panel; left). The sample is extracted from the tube by inclining it and controlling the rate of descent with a piston. The second approach involves doing a second cut to push the sample outwards. A virtue of this approach is that it allows for a more controlled extraction, and it minimizes the risk of the sample getting jammed in the tube. Both options 1 and 2 involve the sample sliding out of the tube and incur the risk of losing the chemical and structural layering of the sample. The third approach involves doing two longitudinal cuts on the side of the tube to expose the whole sample within the tube. It is least likely to disturb the physical integrity of the sample, which stays in place in the tube, but it involves cutting the tube along its length through a white alumina coating (deposited on the tubes to reduce their heat absorption while seating on Mars' surface) possibly using a circular blade (bottom panel; right). The chance of contamination is higher with this third approach as more tube manipulations are involved, more tube material is cut, and the setup to remove or cut the alumina coating will be more involved than the wheel cutter used in approaches 1 and 2.*

**Approach:**

The issue of how to open the tubes was discussed by the team over two telecons. Presentations by engineers Oscar Rendon Perez and Paolo Younse were delivered to explain the design of the tubes and different options for opening them (Fig. 2).

The Rock Sample Team concluded there are three main considerations:

- Need to minimise (and have knowledge of) contamination
- Need to preserve stratigraphy and other textural relationships
- Need to maximise the amount of sample material that ends up in a scientifically useful state from the tubes. For some samples like the detrital sediments or the regolith sample, each small grain may provide a unique record of Mars' surface history, so dust adhering to the tube surface should be recovered to the greatest extent possible. However, such dust will likely represent a small fraction of the total mass and its retrieval could be done later. Or it could be used for quickly surveying the petrography and mineralogy of the core as part of a preliminary examination phase as this material will be of lesser value for other tasks and could be sterilized.

Minimal cutting (*i.e.*, a single radial cut) was considered optimal to minimise potential contamination of trace elements, especially metals, and organic material from the tubes and cutting tools. The structural integrity of the sample would, however, be best preserved with radial and longitudinal cuts; this is considered especially important for sedimentary rocks that may be friable but contain internal stratigraphic structures. The yield may be maximised by at least two radial cuts. These considerations may conflict with each other and the approach to be used will depend on the exact nature of each returned sample. Magnetic contamination should also be minimized during cutting operation and sample handling.

The preferred opening strategies are summarized in Table 1, which ponders each criterion (structure integrity, chemical integrity, and yield) for three categories of samples (consolidated rocks, friable rocks, and loose regolith). We summarize the Rock Team recommendations at the bottom of each column. The rationale for each entry is summarized below:

**Consolidated rocks (example microgabbro).** To minimize the risk of contamination, one radial cut is preferred as cutting by shearing with a hard metal solid wheel will generate little dust, cause little heating, involve no use of fluid, and involve the least amount of tube material of all considered options. To get the sample out of the tube, putting it on a vertical incline and lowering the sample in a controlled manner with a piston would preserve the structural integrity of the sample. One radial cut is likely to preserve the structural integrity of the sample. The cutting wheel will create a metal lip that will protrude in the tube, so provision should be ready to straighten that lip so that the sample can be extracted without rubbing against the lip. With a consolidated sample, there is however a concern that jamming could occur, as a fragment might be trapped in compression between the solid core and the tube wall. A second cut might be needed to push/pull the sample from the other side and free it from such entrapment. Fine dust adhering to the inner tube surface might be difficult to retrieve with a single radial cut. A second radial cut would allow one to get the fine dust out by pushing it out with an appropriate instrument. The Rock Team favours 1 radial cut, with 2 radial cuts possibly needed for sample retrieval in case of jamming and to recover fine dust adhering to the interior tube surface.

**Friable rocks (example detrital sediments).** These rocks are the ones for which preserving the stratigraphy is of upmost importance. The rationale is the same as with consolidated rocks that a single radial cut would be preferred from the point of avoiding contamination. To extract the sample, a single radial cut might be sufficient as the less consolidated nature of those rocks means that they are less likely to be hard jammed in the tube. A possible approach would be to put place a piston against the sample on the opening side with the tube horizontal. The sample tube and piston would then be rotated to a vertical position, and the piston would be lowered in a controlled manner to transfer the sample core in a transparent sample holder (quartz or sapphire) with predesigned longitudinal openings. The reason to transfer the sample vertically is to minimize shear on the tube surface. After vertical transfer of the sample from the tube to the holder, the holder would be rotated back to horizontal to be then opened, giving access to the sample.

Alternatively, it might be possible to 2 radial cuts, and one piston to push the sample out in a slightly inclined orientation and another piston at the open side against the sample to prevent collapse, so the sample keeps its integrity but we can avoid the longitudinal cuts to avoid more risk of contamination. If too friable, the sample could be gently pushed this way into a transparent sample holder with predesigned longitudinal openings, allowing visible inspection of the enclosed protected sample

Letting the sample slide out from one side incurs the risk however that rock fragments will be moved out of sequence, that the sample will disaggregate, and that important chemical features be smeared throughout the core. The latter point could include, for instance, organic distribution. If a layer is highly enriched in organics, sliding the whole sample along the sides may smear the signature throughout the entire core surface. For preserving the stratigraphy, it may therefore be advantageous to make 2 radial cuts and 2 longitudinal cuts to access the core without disturbing it. The constraints on fine dust recovery are the same as with other sample types.

**Regolith.** There is no stratigraphic information to preserve in that sample and little risk of jamming, so a single radial cut is preferred as this minimizes the risk of contamination. The fine dust in the sample may come from afar and each grain will likely tell a story, so complete recovery of dust adhering to the tube inner surface is important.

**Table 1.** Preferred opening strategies depending on rock cohesiveness and criteria considered.

|  | **Consolidated rocks** Example: microgabbro | **Friable rocks** Example: detrital sediments, igneous cumulate rocks | **Regolith** |
| --- | --- | --- | --- |
| Trace element and organic contamination | 1 radial cut | 1 radial cut | 1 radial cut |
| Structural integrity of the sample | 1 radial cut likely OK Maybe 2 radial cuts in case of jamming | 1 radial cut or 2 radial cuts and 2 longitudinal cuts | 1 radial cut |
| Complete retrieval of the sample (including dust) | 1 or 2 radial cuts | 1 or 2 radial cuts | 1 or 2 radial cuts |
| Rock Team recommendation | 1 OR 2 radial cuts | 1 radial cut OR 2 radial cuts and 2 longitudinal cuts | 1 OR 2 radial cuts |

**FINDING:** There is not one single approach for opening the sample tubes that will work sufficiently well for all MSR rock samples. Multiple options need to be available.

**The Rock Sample Team finds that a single approach will not be appropriate for all the rock samples returned**, but instead a flexible and bespoke approach will be needed for each sample tube opening, with all three of the above options available. As a general principle, minimal cutting is favoured as this will also minimise potential contamination issues. However, an overriding consideration is that the structural integrity of the rock sample is key to understanding its petrology, and this should remain intact, even if this requires more processing.

For regolith samples, a single radial cut followed by tipping out the grains is likely to be appropriate, since this will minimise contamination and there is no need to preserve spatial relationships within the tube. For well consolidated (e.g., some igneous rock) samples, a radial cut perhaps followed by a second radial cut may be required to extract the sample completely. For sedimentary rocks, and any friable igneous rocks, the decision is more complex because a longitudinal cut may be necessary to observe and preserve structural relationships, but this must be weighed against potentially contributing more contamination. One possible solution to test for sedimentary samples could be to make one or two radial cuts, then push the sample or let it slide down while keeping its stratigraphy in place (possibly with high inclination to minimize shear along tube surface, with a sliding stopper against the sample to control the sliding rate) into another tube with a closed longitudinal aperture that allows longitudinal opening later.

The physical state of each core (consolidated or friable) will not be known for certain until the samples are bought back to Earth, where CT-scanning will reveal the fine structure of the samples and guide the strategy that adopted for tube opening.

**Future Work:**

The team suggests areas which require more work prior to sample return. These include:

- Investigate how/whether analogue sedimentary samples and aqueously altered cumulate rocks can be removed in a manner that preserves their structural integrity with only one radial cut.
- Investigate ways to efficiently remove the fines left behind after core extraction.
- Impurities in all tube materials, coatings, and opening contraption (e.g., materials used in the saw) must be characterized with appropriate techniques (e.g., ICP-MS). We suggest that a task group be established to undertake an in-depth contaminant characterization campaign.
- Investigate if it is possible to remove the alumina coating without compromising the sample, and without causing damage (e.g. by vibration) to the martian sample inside the core tube.
- Investigate the degree to which the different cutting protocols can introduce contamination.
- Integrate these studies with CT and related scanning techniques.
- Investigate how the cutting and related techniques can be performed in a Biological Hazard Level BSL4 environment.

A concept that is not discussed in this report, but that has been considered elsewhere, is that the opportunity exists to do penetrative imaging/mineralogical characterization of the sample-bearing Mars sample tubes once they make it to Earth, so that we can obtain data on the mechanical state of each sample as received prior to tube opening. This eliminates the need to make guesses based on pre-sampling field data, or accelerations measured by the return spacecraft, etc. That imaging data will give us the opportunity to help make decisions on how to open each tube. We know that for samples with different kinds of mechanical integrity, different tube-opening strategies may be required to avoid the risk of damage that unnecessarily affects the scientific usefulness of the sample.

A component of the technology program is needed to develop the datasets for what happens when tubes containing samples with different degrees of mechanical integrity are opened by each of the three methods described. This will become the basis for future decision-making. We also need data on the real contamination implications of making the horizontal cuts, and what kind of science is affected by such contamination.